\documentclass[a4paper,11pt]{article}

\usepackage{jcappub}
\usepackage[T1]{fontenc} 
\usepackage{graphicx}
\usepackage{graphicx,color}
\usepackage{amssymb}
\usepackage{amsmath}
\usepackage{bm}
\usepackage{subcaption}
\usepackage{hyperref}
\usepackage{multirow}
\usepackage{longtable}
\usepackage{float}
\usepackage{capt-of}
\usepackage[normalem]{ulem}
\restylefloat{table}
\usepackage[dvipsnames]{xcolor}
\usepackage{soul}
\usepackage{verbatim}
\title{\boldmath Sound Speed Resonance in the Gravitational Wave Background as a probe for non-standard early universe cosmologies}

\author[a,b,c]{Igor de O. C. Pedreira,}
\author[a,b]{Amara Ilyas,}
\author[d,e]{Ziwei Wang,}
\author[c]{Leila L. Graef,}
\author[a,b]{Yi-Fu Cai}

\affiliation[a]{Department of Astronomy, School of Physical Sciences,
University of Science and Technology of China, Hefei, Anhui 230026, China}
\affiliation[b]{CAS Key Laboratory for Researches in Galaxies and Cosmology, School of Astronomy and Space Science, University of Science and Technology of China, Hefei, Anhui 230026, China}
\affiliation[c]{Instituto de Fisica, Universidade Federal Fluminense, 24210-346 Niteroi, RJ, Brazil}
\affiliation[d]{Department of Strategic and Advanced Interdisciplinary Research, Pengcheng Laboratory, Shenzhen, Guangdong 518066, China}
\affiliation[e]{Key Laboratory of Dark Matter and Space Astronomy, Purple Mountain Observatory, Chinese Academy of Sciences, Nanjing 210033, China}

\emailAdd{igorpedreira@id.uff.br}
\emailAdd{aarks@ustc.edu.cn}
\emailAdd{wangzw@pcl.ac.cn}
\emailAdd{leilagraef@id.uff.br}
\emailAdd{yifucai@ustc.edu.cn}

\abstract{Gravitational waves constitute a powerful probe of the underlying theory of gravity. In extensions of general relativity, additional degrees of freedom, such as scalar fields in the gravitational sector, can modify their propagation through changes in the effective friction term and propagation speed. These modifications may potentially induce  resonant phenomena leading  to distinctive signatures in the gravitational wave spectrum. One important aspect to be investigated is whether the resonances can be strong enough to enhance the underlying background of primordial tensor modes to levels detectable by upcoming gravitational wave detectors, such as LISA or the Einstein telescope. The characteristic peaks in the SBGW spectrum depend on the parameters of the resonant model as well as on the parameters of the primordial tensor spectrum, such as $r$ and $n_{t}$. Thus these resonance effects open a powerful pathway to explore physics of the very early Universe by amplifying otherwise feeble signals to experimentally detectable levels. Here we analyze how the signals of the primordial Universe can resonate in these scenarios, bringing the early universe physics into the realm of experimental access.}

\begin{document}

\maketitle
\flushbottom

\section{Introduction}
\label{sec:intro}
Research on Gravitational Waves (GW) has grown in the past decade, since the first detection from astrophysical sources by LIGO \cite{LIGOScientific:2016aoc}. Since then, the community has intensified its efforts to detect gravitational wave signals across all frequencies. Later in 2023 strong evidence on the presence of a stochastic signal in the Pulsar Timing Arrays data was reported by NANOGrav \cite{NANOGrav:2023gor}, and later, similar levels of evidence were reported by several other PTA collaborations \cite{Goncharov:2021oub, EPTA:2021crs, Antoniadis:2022pcn}. Besides astrophysical sources, cosmological phenomena can also generate a stochastic background of gravitational waves (SGWB). Among the cosmological sources \cite{NANOGrav:2023hvm}, we can mention the quantum fluctuations generated in the very early universe, which are stretched out from the Hubble radius, and later, when re-entering the Hubble horizon, behave like relic gravitational waves, forming a stochastic background. 

Compared to the astrophysical signals, cosmological sources can predict different shapes for the SGWB power spectrum, which can be probed across all frequency ranges. The power spectrum of the primordial GW can be parametrized by the tensor-to-scalar ratio $r$, constrained by the PLANCK collaboration \cite{Planck:2018vyg, Tristram:2020wbi}, and the tensor spectral index $n_t$. There exists a wide variety of models in the literature predicting either a blue-tilted spectrum ($n_t > 0$) or a red-tilted spectrum ($n_t < 0$). However, no direct observational constraint on the tensor spectral index is currently available \cite{Kuroyanagi:2014nba, Vagnozzi:2023lwo, Kawai:2023nqs}. 
 
Models predicting a blue tilt are of particular interest, since they enhance the gravitational wave amplitude at higher frequencies, potentially bringing the signal within the sensitivity range of current or future GW detectors \cite{Mishra:2025nnu}. Additional mechanisms can also predict spectra with a signal  that grows with the frequency across different frequency bands. These include, for instance: non-canonical inflationary models \cite{Piao:2004tq, Kobayashi:2011nu, Cai:2014bda, Cai:2014uka, Cannone:2014uqa, Dinda:2014zta, Wang:2014kqa, Papanikolaou:2022did, Choudhury:2023kam, Jiang:2023gfe, Datta:2023vbs, Datta:2023xpr, Chianese:2024nyw, Athron:2024fcj},  non-instantaneous reheating \cite{Vagnozzi:2023lwo, Ben-Dayan:2023lwd, Li:2025udi}, natural inflation coupled to gauge fields \cite{Barnaby:2011qe, Domcke:2016bkh, Jimenez:2017cdr, Papageorgiou:2019ecb, DEramo:2019tit, Papanikolaou:2024cwr}, bounce models \cite{Peter:2006id, Cai:2012va,Cai:2014bea, Battefeld:2014uga, Brandenberger:2016vhg, Cai:2016hea, Odintsov:2020vjb, Papanikolaou:2024fzf}, cosmological first-order transitions \cite{Liu:2015psa, Santos:2022hlx, Ahmed:2023pjl, Caprini:2024hue, Schmitz:2024gds}, particle production \cite{Cook:2011hg, Mukohyama:2014gba, Hipolito-Ricaldi:2016kqq}, axion couplings \cite{Pajer:2013fsa, Dimastrogiovanni:2016fuu, Obata:2016oym, Adshead:2016omu, Iacconi:2020yxn, Kume:2025lvz, vonEckardstein:2025oic} departures from GR and modified gravity theories \cite{Myrzakulov:2015qaa, Fujita:2018ehq, Odintsov:2019evb, Odintsov:2020nwm, Nojiri:2020pqr, Oikonomou:2022yle, Oikonomou:2022ijs, Capozziello:2024vix, Oikonomou:2024etl, Capozziello:2025htb, Papanikolaou:2025ddc, Li:2024fxy}. 

The GW detections nowadays happen mainly in earth-based interferometers, such as the LIGO-VIRGO-KAGRA collaboration \cite{LIGOScientific:2014pky, VIRGO:2014yos, KAGRA:2018plz} in the $10^0 - 10^4 \mathrm{Hz}$ frequency range. A future earth-based detector that will considerably improve the current detector's sensitivity in the high frequency range  will be the Einstein Telescope \cite{Punturo:2010zz, Hild:2010id, ET:2025xjr} an underground interferometer operating in the $10^{-3} - 10 \mathrm{Hz}$ frequency range planned for later in the 2020's. In a lower frequency range $10^{-5} - 10^{-1} \mathrm{Hz}$, the LISA space interferometer is planned to be launched in the next decade \cite{LISA:2017pwj, Barausse:2020rsu, LISA:2022kgy, LISACosmologyWorkingGroup:2024hsc}. 
In even lower frequencies, Pulsar Timing Arrays (PTA) may, in the near future, allow to probe the SGWB signal in the nanohertz frequency band \cite{NANOGrav:2023gor, NANOGrav:2023tcn, NANOGrav:2023hvm, NANOGrav:2023hde, Xu:2023wog, EPTA:2023fyk, EPTA:2023sfo, EPTA:2023gyr, Zic:2023gta, Reardon:2023gzh, Antoniadis:2022pcn}.  Probing a similar frequency range, the Square Kilometer Array (SKA) \cite{Weltman:2018zrl}, shall soon start operating increasing considerably the sensitivity in that band. 

So far the Planck satellite has imposed an upper limit on the tensor-to-scalar ratio to be $r<0.035$ \cite{Planck:2018jri} \footnote{Tighter upper limits on $r$ can be obtained by combining Planck data with other cosmological probes, as shown, for instance, in Ref.~\cite{Tristram:2021tvh, Balkenhol:2025wms}. These constraints are of the same order as the Planck-only bounds, which were adopted in the present analysis.}. 
Other CMB experiments, such as SPT-3G \cite{SPT-3G:2014dbx, SPT-3G:2022hvq, SPT-3G:2025vtb, SPT-3G:2025bzu} and especially the LITEBIRD satellite \cite{Hazumi:2019lys, LiteBIRD:2022cnt, Anand:2025acx}, which is being planned to search for these B-modes, intend to measure the tensor-to-scalar ratio up to a level of $10^{-3}$ (for a review, check the reference~\cite{Komatsu:2022nvu}). In the literature, there have been various models to study a decrease in the value of $r$ \cite{Kallosh:2013hoa, Cai:2014bda, Zhu:2021whu, Rodrigues-da-Silva:2021jab, Shtanov:2022pdx, Ferreira:2025lrd, Kallosh:2025sji}. 
In addition to CMB constraints, an important constraint on the  SBGW spectrum comes from the Big Bang Nucleosynthesis (BBN) bound on the relativistic degrees of freedom on the primordial universe, which puts an upper limit on the amplitude of the GW spectrum across a long range of frequencies \cite{Kuroyanagi:2014nba, Cyburt:2015mya, Benetti:2021uea, Wang:2025qpj}. 

The search for SBGW can be of paramount importance not only to probe the physics of inflation but also as a test for possible departures of general relativity (GR) \cite{DeLeo:2025lmx, Balaudo:2023klo, Maselli:2016ekw}. Modified gravity theories arises as a interesting alternatives to GR, offering possible explanations for several outstanding problems in cosmology. These include the origin of the current   accelerated expansion \cite{Cognola:2006eg, Langlois:2015cwa, Drepanou:2021jiv, Brax:2021wcv, Oikonomou:2022wuk, Trivedi:2024inb, Lin:2025gne, Feleppa:2025clx}, explain the evolution of the primordial universe \cite{Boulware:1985wk, Ferraro:2006jd, Mandal:2018bnr, Zhang:2021ppy}, or even the behavior and nature of dark matter \cite{Mocz:2019pyf, Hui:2021tkt, Choi:2021aze, Tsai:2023zza}. In many cases, such theories are constructed by introducing additional degrees of freedom, most commonly a scalar field, alongside the usual tensor modes of GR. These frameworks are generally referred to in the literature as scalar–tensor theories \cite{Langlois:2015cwa}. 

The dynamics of  scalar fields in modified gravity theories under certain conditions may cause an amplification of the gravitational wave spectrum. Time-oscillations of such fields at the bottom of their potential may lead to an effect called parametric resonance, which results in a exponential enhancement of GW power, leaving significant imprints in different frequencies of the GW spectrum \cite{ Guzzetti:2016mkm, Cai:2021yvq, Domenech:2024drm}. This effect is also called sound speed resonance (SSR), because those resonances can cause changes in the propagation speed of the sound waves due to the scalar field, and has previously been studied in the literature \cite{Cai:2018tuh, Cai:2019jah, Chen:2019zza, Chen:2020uhe, Addazi:2022ukh, Jin:2023wri, Wu:2024deu, Brahma:2026fre}.

In recent years, the possibility of the existence of ultra light scalar fields non minimally coupled to gravity started to get more attention in community, because such a field can be considered as a possible candidate for constituting a class of dark matter named ultra light dark matter (ULDM). These kind of particles can arise from axion models\cite{Hu:2000ke, Arvanitaki:2014faa, Marsh:2015xka, Lague:2023wes}, dark photons \cite{Nelson:2011sf, Arias:2012az, Jaeckel:2021xyo, Brahma:2025vdr} or light particles generated in string theory models \cite{Svrcek:2006yi, Arvanitaki:2009fg, Cicoli:2012sz, An:2020jmf, Lague:2021frh, Sheridan:2024vtt, Capanelli:2025ykg}. 
Those particles are proposed as a way to address the behavior of dark matter in small scales, maintaining the large scale behavior intact.
Such particles can exhibit a wave-like nature, thanks to its low mass $m \sim 10^{-22}-10^{0} \mathrm{eV}$, which can help resisting the collapse of dark matter halos at galaxy scales and suppress the formation of small mass halos \cite{Hui:2016ltb, Ferreira:2020fam, Hui:2021tkt, Delgado:2023psl, Matos:2023usa, Lu:2024uyv, OHare:2024nmr, Eberhardt:2025caq, Yang:2025vcb, Brandenberger:2025gks}.

ULDM provides a well-motivated example of a coherently oscillating scalar field. When such a scalar is non-minimally coupled to gravity, its oscillations can induce periodic time dependence in the propagation of gravitational waves, leading to the parametric resonance phenomena previously mentioned.
In this paper we will consider an ULDM field in the context of scalar tensor theories with higher order derivatives, given by quadratic Degenerate Higher Order Scalar Tensor (DHOST) Theories \cite{Langlois:2015cwa, Langlois:2015skt, Crisostomi:2016czh, BenAchour:2016cay, Kobayashi:2019hrl}, which avoid the presence of the so called Ostrogradski instability that plagues general scalar tensor formulations \cite{Woodard:2006nt, Woodard:2015zca}. 
In this class of theories, the amplification of the GW power spectrum can happen due to a change in the propagation speed of the gravitational waves, in the friction term or due to  modifications on the background evolution which alters the Friedmann Equation \cite{Pettorino:2014bka, Xu:2014uba, Cai:2023ykr}. 
We will 
consider disformal transformations to work in the frame where tensor modes propagate as in GR \cite{Bekenstein:1992pj,  Creminelli:2014wna, Gleyzes:2014qga}. In this frame, parametric resonances arising from the oscillations in the GW propagation speed can be associated to oscillations in the scale factor of the GW frame, which help us find potential cancellation effects.
In order to simplify our calculations, we will consider an Effective Field Theory (EFT) approach for DHOST and require that the modifications are below the cut-off scale of the EFT \cite{Ilyas:2020qja, Cai:2023ykr}. 
We consider the DHOST scenario in which the stochastic gravitational wave background evolves through a standard radiation-dominated universe, producing a smooth baseline spectrum as predicted by GR \cite{Cai:2023ykr}. We will consider that certain modes experience resonant amplification due to time-dependent modifications in the gravitational-wave propagation on this smooth baseline. Most other modes continue to follow the standard evolution, resulting in narrow, localized bands that stand out on the background spectrum. 

Our interest in this work is to test whether the resonance bands can be used as a way to probe a non-standard background evolution of the early universe. In principle, the resonance effect can happen in different periods of the evolution of the universe, but we focus on it happening during radiation domination era. We will explore how the amplitude of these resonant bands is affected by the underlying primordial tensor spectrum. We assume the primordial tensor spectrum to follow the standard parametrization, without supposing that modified gravity theory affects the generation of the tensor modes.
We will remain agnostic on the underlying early universe scenario, whose spectrum we  simply parametrize by  different combinations of the tensor-to-scalar ratio $r$ and the tilt of the tensor spectrum $n_t$. This approach allows us to explore a broad range of scenarios, with particular emphasis on blue-tilted primordial spectra, as these can enhance the gravitational-wave amplitude at frequencies within the sensitivity band of LISA. 
Specifically, we are going to analyze how the SSR effect caused by the dynamics of ULDM oscillations in a quadratic DHOST theory can affect the detection prospects of the SGWB in different early universe scenarios. While the quantitative analysis here will be carried out for ULDM scalar field within the framework of DHOST scalar tensor theories, we argue that the underlying mechanism is more general.  
Despite the quantitative results remains  model-dependent, any consistent modified gravity theory in which an oscillating scalar field directly modulates the GW propagation coefficients can, in principle, exhibit similar resonant effects. DHOST theories provide a controlled and observationally viable setting, free of instabilities and compatible with gravitational-wave constraints, but the qualitative conclusions extend to broader classes of scalar tensor theories. 

This paper is organized as follows: In section 2, we present the derivation of the expression for   the GW spectral energy density and we comment about the astrophysics foregrounds and the BBN limits.  
In section 3, we describe the ULDM scenario that predicts the parametric resonance effect.
In Section 4, we present our results and demonstrate how the resonance reshapes the region of parameter space accessible to LISA in the $r$-$n_T$ plane. 
Finally, we  give our final remarks in section 5.

\section{The Spectrum of the Gravitational Waves} \label{sec:2}
In this section, we review the derivation of the stochastic gravitational wave background spectrum arising from tensor perturbations of the Friedmann-Lemaitre-Robertson-Walker (FLRW) metric.

\subsection{Primordial Gravitational Waves}
If we consider small perturbations ($\lvert{h_{ij}} \rvert \ll 1$) over an isotropic and homogeneous FLRW metric in the transverse traceless gauge (\(\partial_i h_{ij} = h_{ii} =0\)), at first order, one can write the line element as \cite{Caprini:2018mtu}:

\begin{equation}\label{eq_2.1}
    ds^2 = a^2(\eta) \left[- d\eta^2 + (\delta_{ij} +h_{ij}) dx^i dx^j \right] \, .
\end{equation}

By using this metric in the Einstein equation, we obtain a second order differential equation for the evolution of the perturbation $h_{ij}$. In Fourier space, we can write for both polarization modes \cite{Vagnozzi:2023lwo}:

\begin{equation} \label{eq_2.2}
    {h_{k}''}^\lambda + 2\frac{a'}{a} {h_{k}'}^\lambda + k^2 {h_{k}}^\lambda = 0   \, ,
\end{equation}
where the prime $'$ denotes derivative with respect to conformal time \(d/d\eta\). With the solution of this equation, we can write down the gravitational wave spectrum as:

\begin{equation}
    \Omega_{GW} (k, \eta) = \frac{1}{\rho_c (\tau_0)} \frac{\partial \rho_{GW} (k, \tau)}{\partial \ln{k}} \, .
\end{equation}
where $\rho_c = H_0^2/8 \pi G$ is the critical density of the universe today and $\rho_{GW}$ is related to the perturbation coefficient as \cite{Watanabe:2006qe}:
\begin{equation} \label{eq_2.4}
    \rho_{GW} (k, \eta) = \frac{\langle h'_{ab} h'^{ab} \rangle}{32 \pi a^2 G}
\end{equation}

We can rewrite the numerator on the right side of eq.~\eqref{eq_2.4} in terms of the power spectrum, defined as: \( \mathcal{P}_T (\eta,k) = {2k^3}/{2 \pi^2} \langle h_{ij} h^{ij} \rangle \). We can also rewrite the primordial power spectrum in terms of the transfer function $\mathcal{T}(\eta, k)$ as:
\begin{equation}
    \mathcal{P}_T(\eta,k) \equiv \mathcal{P}^{prim} (k) \times \left[ \mathcal{T}(\eta, k) \right]^2 \, .
\end{equation}

$\mathcal{P}^{prim} (k) $ is related to the modes that left the horizon during the inflation and can be parametrized as a power law around the CMB pivot scale, $k_\star = 0.05Mpc^{-1}$ \cite{DEramo:2019tit}:
\begin{equation} \label{eq_2.6}
    \mathcal{P}^{prim} (k) \equiv A_T \left(\frac{k}{k_\star}\right)^{n_t}  \, ,
\end{equation}
where $A_T$ is the amplitude of the primordial tensor perturbations, which can be written in terms of the amplitude of scalar perturbations $A_S$ and the tensor-to-scalar ratio $r$: $A_t = r \times A_S$.
The transfer function describes the subsequent time evolution of the perturbation modes, and is defined as \( \mathcal{T}(\eta, k) \equiv \frac{h_k(\eta)}{h_k^{prim}} \). By using all the previous equations, we can write the expression for the GW power spectrum as:
\begin{equation} \label{eq_2.7}
    \Omega_{GW} (k, \eta) = \frac{1}{12 a^2(\eta) H^2 (\eta)} \left[\frac{d}{d \eta}\mathcal{P}_T(\eta,k) \right]^2 = \frac{\mathcal{P}^{prim} (k)}{12 a^2(\eta) H^2 (\eta)} \left[\mathcal{T}'(\eta, k) \right]^2 \, .
\end{equation}

Although primordial power spectrum of gravitational wave generated by slow roll inflation has to obey the consistency relation \cite{Planck:2015fie, Guzzetti:2016mkm}, in which the tensor-to-scalar ratio $r$ and the spectral index $n_t$ are related via: $n_t = -\frac{r}{8}$, indicating a slightly red spectrum, there are early universe scenarios predicting blue tilted spectrum of gravitational wave. For example:
\begin{itemize}
    \item Fields satisfying the violation of the Null Energy Condition (NEC) during inflation, such as G-inflation \cite{Kobayashi:2010cm}, break the consistency relation, which can produce a blue tilted tensor perturbation. 
    \item If the universe underwent a distinct kinetic or geometric phase prior to the standard inflationary expansion, the initial vacuum state of the perturbations can be altered into a non-Bunch-Davies state (often parameterized as $\alpha$-vacua).\footnote{Historically, such modified initial states were sometimes invoked to model effective trans-Planckian physics \cite{Martin:2000xs}. However, the Trans-Planckian Censorship Conjecture (TCC) posits that in a consistent quantum theory of gravity, modes that originate at sub-Planckian scales are strictly forbidden from exiting the Hubble radius and becoming macroscopic \cite{Bedroya:2019snp,Bedroya:2019tba}.} The mode mixing—described by Bogoliubov transformations between the standard Bunch-Davies vacuum and the actual state—modifies the amplitude and scale dependence of the power spectra. Depending on the precise Bogoliubov coefficients generated by the pre-inflationary transition, this can introduce a scale-dependent modifications that tilts the tensor spectrum blue \cite{Mohanty:2014kwa}. 
\end{itemize}

The scenarios mentioned above predict $ n_t >0$ and break the consistency relation between $n_t$ and $r$, respectively. In this work, rather than discuss early universe scenarios in detail, we parameterize the primordial tensor perturbations with free parameters $r$ and $n_t$ not supposing any consistency relation and study the enhancement of gravitational waves through parametric resonance phenomenologically.

\subsubsection{Solution for the perturbations modes}

The evolution of the modes $h_k$ is very complicated and cannot be solved analytically for a random expansion history \cite{Maggiore:1999vm, Watanabe:2006qe}. But for constant equation of state $\omega$, it has been obtained before in the literature for different scenarios \cite{Soman:2024zor}.

The gravitational waves we measure today are supposed to be originated by quantum fluctuations in the primordial universe. As the universe undergoes accelerated expansion during inflation, those modes are stretched to super-horizon scales ($k <aH$) and freeze, which leads to \(h_k^{prim} \propto \mathrm{Const.}\) \cite{Saikawa:2018rcs}. When they re-enter the horizon they  evolve classically, according to $h_k (\tau) = h_k^{prim} \times \mathcal{T}(\eta, k)$. 

When the modes are deep inside the horizon, as the case that will be considered in this paper, we can approximate the solution for the evolution of the modes considering the WKB approximation \cite{Kuroyanagi:2008ye, Saikawa:2018rcs}. Considering that, the modes can be approximate as:
\begin{equation}
    h_k(\eta) \propto \frac{1}{a(\eta)} e^{\pm ik\eta} \, .
\end{equation}

With it, it's possible to rewrite the transfer function as the following expression:
\begin{equation} \label{eq_2.9}
    \mathcal{T} (k \eta)= \frac{a(\eta_{hc})}{a(\eta)} e^{\pm ik \eta} \, . 
\end{equation}

To construct the spectrum, we will need the derivative of the transfer function squared. By considering Eq.~\eqref{eq_2.9}, and using the fact that $k = a_{hc} H_{hc}$ during the reentry, we can rewrite \cite{Saikawa:2018rcs}:
\begin{equation} \label{eq_2.10}
    \left[\mathcal{T}'(k\eta)\right]^2 \approx \frac{k^2}{2} \left(\frac{a(\eta_{hc})}{a(\eta)}\right)^2 = \frac{a^4(\eta_{hc}) H^2(\eta_{hc})}{2a^2(\eta)} \, .
\end{equation}

Applying eq.~\eqref{eq_2.10} to eq.~\eqref{eq_2.7}, we obtain the following expression for the spectrum:
\begin{equation}
    \Omega_{GW} (k, \eta) = \frac{\mathcal{P}^{prim} (k)}{24} \frac{a^4(\eta_{hc}) H^2(\eta_{hc})}{a^4(\eta) H^2 (\eta)} \, .
\end{equation}

We can rewrite this expression by considering that during the horizon crossing $ H^{2} = \Omega_\omega H_0^2 a^{-3(1+\omega)}$, where $\Omega_\omega$ is the density parameter of the dominant component of given epoch, and $k = a_{hc}H_{hc}$. Therefore,
\begin{equation}
    k \propto a_{hc}^{-\frac{1+3\omega}{2}} \, .
\end{equation}
Using this expression we obtain for the spectrum today,
\begin{equation}
    \Omega_{\mathrm{GW}}(k, \eta_{0}) = \frac{\mathcal{P}_{prim}(k) \Omega_\omega}{24a_{0}^{4}H_{0}^{2}}k^{2\big(\frac{\omega - 1/3}{\omega + 1/3}\big)} \, .
    \end{equation}
    
By using the parametrization given by Eq.~\eqref{eq_2.6} and using the relation between wave-number and frequency:
\begin{equation} \label{eq_2.14}
    f = \frac{1}{2\pi} \frac{k}{a_0} \, ,
\end{equation}
we can obtain the final expression for the spectrum of the modes that re-enter the horizon during radiation-dominated (\(\omega = 1/3\)) era as\footnote{By eq.~\eqref{eq_2.14}, the CMB pivot scale $k_\star$ corresponds to a frequency of $f_\star = 7.5 \cdot 10^{-15} \mathrm{Hz}$.} \cite{Mishra:2021wkm, Kite:2021yoe}:
\begin{equation} \label{eq_2.15}
    \Omega_{GW} (f, \eta) = \frac{\Omega_{rad}}{24} r \cdot A_S \left(\frac{f}{f_\star}\right)^{n_t} \, .
\end{equation}

\subsection{Astrophysical Foreground} \label{sec:Astro}
A stochastic signal generated by the population of binary systems scattered around our galaxy could be measured by the current and next generation GW detectors. This signal usually comes from "unresolved" sources, that do not come from a single event, but composes a stochastic background \cite{Allen:1996vm}. If that signal is not strong enough, it could go undetected or could be indistinguishable from noise. It is crucial to understand the expected astrophysical signal across the frequencies bands probed by the detectors. 

The main sources of astrophysical foreground in different frequency ranges are: the unresolved populations of Galactic and Extragalactic White Dwarfs binaries (GWD and EGWD, respectively); the unresolved stellar mass binaries of Black Holes (BBH), Neutron Stars (BNS) and Black Hole- Neutron Star binary (BH-NS); the contribution due to the coalescence of Massive Black Hole Binaries (MBHB). It is possible to make approximate analytical fits for the astrophysical foreground's contributions \cite{Campeti:2020xwn}. 

Inside the LISA detection band, the first three astrophysical systems give the major contribution to the astrophysical foreground signal. Usually, those contributions are represented  in terms of the spectral density $S_h(f)$ \cite{Maggiore:1999vm, Moore:2014lga}. By writing the contributions in terms of the spectral density, the comparison with the signal-to-noise ratio becomes more transparent. The spectral density is related with the GW spectrum through the equation\cite{Mingarelli:2019mvk},
\begin{equation} \label{eq_2.16}
    \Omega_{GW}(f) = \frac{4\pi^2}{3H_0^2} f^3 S_{h}(f) \, .
\end{equation}

The expressions for the spectral density of different astrophysical sources can be found, for instance in Ref.~\cite{Campeti:2020xwn}.
{In particular,  the astrophysical foregrounds considered in this work  include the contributions from unresolved galactic and extragalactic white dwarf binaries, as well as from compact binary mergers (binary neutron stars, binary black holes, and black hole - neutron star systems).  We refer the reader to Refs.~\cite{Campeti:2020xwn, Cornish:2017vip, Robson:2018ifk, Nishizawa:2011eq, LIGOScientific:2025bgj} for the detailed expressions.}



\begin{figure} [h]
    \centering
    \includegraphics[width=0.6\linewidth]{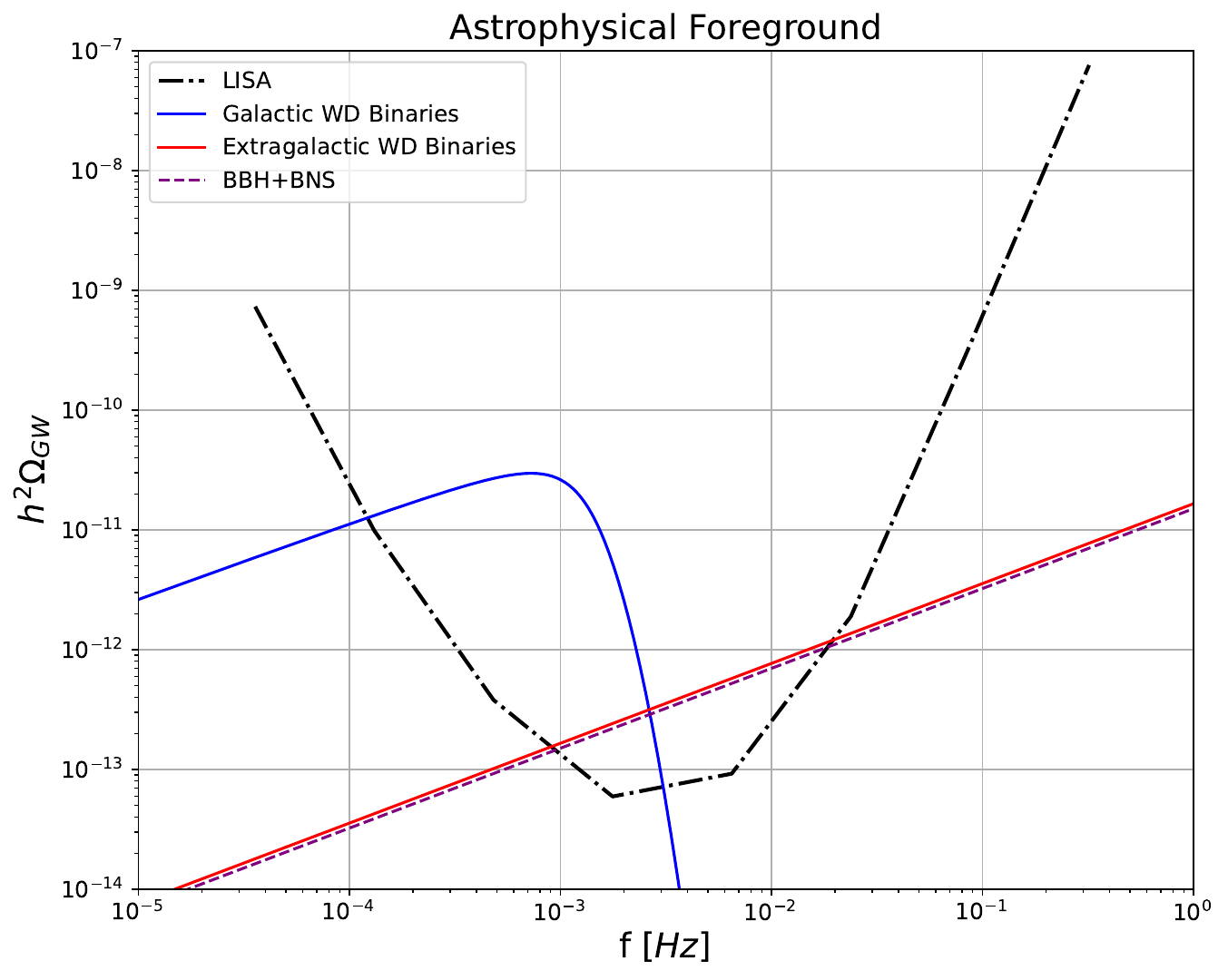}
    \caption{SGWB spectrum for the astrophysical foreground. We shown the unresolved Galactic WD in blue, the unresolved Extragalactic WD in red and the sum of the BBH and BNS in magenta. In Black, we show the LISA's sensitivity curve.}
    \label{fig:1}
\end{figure}

In figure \ref{fig:1}, we can see the astrophysical contributions for the stochastic GW spectrum that can contribute to a detectable a signal inside the LISA frequency band. 


\subsection{Big Bang Nucleosynthesis Bound}

In principle, the GW can be generated with any amplitude, depending on the physics that creates those modes and how they evolve, as we can see from Eq.~\eqref{eq_2.15}. Although, there are a few limits on the spectrum arising from the observations. First, we have a detection upper bound from the absence of a cosmological signal in higher frequencies from the LIGO/VIRGO/KAGRA detectors of the order \(\Omega_\mathrm{GW} \leq 6.0 \times 10^{-8}\) in the reference frequency $f_{LV} = 25\mathrm{Hz}$ \cite{Kuroyanagi:2020sfw}. On the other hand, in the lowest frequencies we have constraints imposed by the  CMB experiments.

In addition, the big bang nucleosynthesis  implies in a constraint to a broad frequency range of the spectrum. Primordial gravitational waves contributes to the total energy density of the extra relativistic species during the early universe, which changes the expansion rate and affects the abundance of the light elements. In order to not violate the BBN bound, the total energy density of the gravitational waves should obey \cite{Kuroyanagi:2020sfw}:
\begin{equation} \label{eq_2.17}
    \int_{f_{BBN}}^{f_{end}} d(\ln f) \, \Omega_{GW} (f) h^2\leq 5.6\cdot 10^{-6} \left( N_{eff}^{(Upper)} - 3.046 \right) \, ,
\end{equation}
where $N_{eff}^{(Upper)}$ is the upper bound on the effective number of relativistic degrees of freedom, taken to be $\Delta N_{eff} = N_{eff}^{(Upper)} - 3.046 \sim 0.300$ \cite{CosmoVerseNetwork:2025alb}, the lower limit in the integral is taken to be frequency corresponding to the mode entering the horizon at the time of BBN $f_{BBN} = 10^{-10}\mathrm{Hz}$ and $f_{end} = 10^7\mathrm{Hz}$ is set to be the scale on the end of inflation, which correspond to a temperature around $10^{15} \mathrm{GeV}$ \cite{Kuroyanagi:2020sfw, Benetti:2021uea}. {Unless the GW spectrum has a very narrow peak with width \(\Delta f \ll f\), the BBN bound can be interpreted as a maximum bound on the amplitude of the GW spectrum \cite{Caprini:2018mtu}.}

\section{Parametric Resonance in DHOST} \label{sec:3}
In the previous section, we discussed the spectrum of the SGWB from cosmological origin and the  astrophysical foreground  expected within the LISA detection band. However, it remains to be studied how parametric resonance affects this background. In this section, we describe the ULDM formalism and its influence on the gravitational wave background.

\subsection{ULDM scenario}
Let us consider a ULDM setup where a parametric resonance of the Gravitational Waves occurs due to modifications in the background during the radiation dominated era. For that, let's apply the DHOST formalism developed in Ref.~\cite{Cai:2023ykr} in a scenario where the ULDM is non-minimally coupled to gravity. We considering the following action \cite{Kobayashi:2019hrl, Gandolfi:2023hwx}:
\begin{equation}
    S = \int d^4 x\sqrt{-\tilde{g}} \left[ \frac{1}{2}\tilde{R} + \frac{1}{2}\tilde{X} + \frac{1}{2}m \phi^2 - \frac{\lambda}{4} \phi^4 + \tilde{G}^{\mu\nu}\frac{\phi_\mu \phi_\nu}{\Lambda^2} \right] + S_M(\tilde{g}_{\mu\nu} ,\psi) \, .
\end{equation}

We work in the so-called GW frame, in which the propagation of GW satisfies the equation of motion of canonical tensor modes given by:
\begin{equation}
    v_k'' + \left(k^2 - \frac{a''}{a}\right) v_k =0\,.
\end{equation}

In GW frame, the modifications of gravity are encoded in the scale factor. It is convenient to rewrite the DHOST theory in the GW frame by considering the following general disformal transformation \cite{Cai:2023ykr}:
\begin{equation}
    \tilde g_{\mu\nu} = \left(1+\hat D(\phi,X)X\right)C(\phi,X)g_{\mu\nu},\,
\end{equation}
where $C(\phi,X)$ and $\hat D(\phi,X)$ are expanded up to $\mathcal O(\frac{X}{\Lambda^2})$ as 
\begin{align}
     C(X,\phi) & \simeq  1 - \frac{6X}{\Lambda^2},\,\\
     \hat D(X,\phi) & \simeq  - \frac{2}{\Lambda^2} + \frac{8X}{\Lambda^4},\,
\end{align}
in the perspective of EFT of gravity with a cut-off scale $\Lambda$. 

Considering the background equation of motion for $\phi$ up to first order of $\mathcal O(\frac{X}{\Lambda^2})$:
\begin{equation} \label{eq_3.27}
    \ddot{\phi} + 3H\dot{\phi} + m^2 \phi + \lambda\phi^3 = \mathcal{O}\left(\frac{X}{\Lambda^2} \right) \, .
\end{equation}

In the regime where the self-interaction term dominates the potential $\lambda\phi^2\gg m^2$, as expected to happen in the early universe, we obtain:
\begin{equation} \label{eq_3.7}
    \phi \simeq \frac{\phi_0}{a} \mathrm{sn}(x,-1) \quad, \quad X \simeq \frac{\phi_0^2\omega^2}{2a^4}\left(\mathrm{cn}(x,-1) \mathrm{dn}(x,-1) - \frac{\mathcal{H}}{\omega} \mathrm{sn}(x,-1)\right)^2 \, ,
\end{equation}
where
\begin{equation} \label{eq_3.29}
    x = \omega(\tau-\tau_\star) \quad , \quad \omega = \sqrt{\lambda \phi_0^2/2} \, .
\end{equation}

In the previous equations, $\mathrm{sn}(x,-1)$, $\mathrm{cn}(x,-1)$ and $\mathrm{dn}(x,-1)$ are the Jacobi elliptic functions, $\tau_\star$ is the conformal time at the onset of the resonance and $\omega$ is the oscillation frequency of the ULDM field when it dominates the potential energy. From now on, we will use the convention $\mathrm{sn}(x, -1) \equiv \mathrm{sn}(x),\,\mathrm{cn}(x, -1) \equiv \mathrm{cn}(x),\,\mathrm{dn}(x, -1) \equiv \mathrm{dn}(x)$, for the three functions separately.

When the self-interaction term dominates, this scalar field behaves like radiation. Only when the mass term dominates the potential, it starts to behave like dark matter, which happens for a time \(\tau \geq \tau_t\), where \(\tau_t\) is the transition time, defined through the relation $m^2 a(\tau_t)^2 = \lambda\phi_0^2$. We require this to occur before the radiation-matter equality, \(\tau_t < \tau_{eq}\) \cite{Cai:2023ykr}.

For the tensor modes, we will focus on the dominant terms of Eq.~\eqref{eq_3.7} when \(\omega\gg\mathcal{H}\) and \(\phi_0 \ll 1\). Expanding the scale factor and its derivative over conformal time in the GW frame, up to the order of $\beta = \mathcal{H}/\omega$, we obtain the following equation for canonical tensor mode variable,
\begin{multline} \label{eq_3.9}
\frac{\mathrm{d}^2 v_k}{\mathrm{d} x^2} + \bigg[ \kappa^2 - 120\alpha^2 \big(\mathrm{sn}^2(x) - 4\mathrm{sn}^6(x) + 4\mathrm{sn}^{10}(x)\big) \\
+ 120\beta\alpha^2 \mathrm{cn}(x)\mathrm{sn}^3(x)\mathrm{dn}(x)\big(8 - 11\mathrm{sn}^4(x)\big) \bigg] v_k \simeq 0 \, ,
\end{multline} where the parameters are defined as:
\begin{equation} \label{eq_3.10}
\kappa = \frac{k}{\omega}, \quad \alpha = \frac{\phi_0^2 \omega^2}{a^4 \Lambda^2} \equiv \frac{\alpha_\star}{(1 + \beta_\star x)^4}, \quad \beta = \frac{\mathcal{H}}{\omega} \equiv \frac{\beta_\star}{1 + \beta_\star x}, \quad a(x) = a_\star \left(1 + \beta_\star x\right) \,.
\end{equation}
\
In Eq.~\eqref{eq_3.10}, the terms with a star means that it was evaluated at \(\tau=\tau_\star\). The leading term in Eq.~\eqref{eq_3.9} scales as $\alpha_\star$, this term controls the strength of the resonance, and corresponds to \(X^2/\Lambda^4\). The parameter \(\beta_\star\) determines how long the mode will remain inside the resonance band.

\subsubsection{Key Parameters}
Now that we have described the formalism, we can study how the parameters affects the resonance, which will be important to describe the results in Sec.~\ref{sec:4}. First, as stated in Ref.~\cite{Cai:2023ykr}, the enhancement is highly sensitive to \(\alpha_\star\), and for it to be significant, we need to require that:
\begin{equation}
    \frac{\alpha_\star^4}{\beta_\star} \sim \left(\frac{H_\star}{\Lambda}\right)^8 \left(\frac{a_t}{a_{eq}}\right)^4 \frac{m}{H_\star}
    \frac{\alpha(\tau_t)}{a_\star}\sim
    10^{21} \left(\frac{m}{\mathrm{eV}}\right)
    \left(\frac{\mathrm{MeV}}{T_\star}\right)
    \left(\frac{T_{eq}}{T_t}\right)^5
    \left(\frac{H_\star}{\Lambda}\right)^8 \gtrsim 1 \, .
\end{equation}

The previous equation provides a lower bound for the mass of the scalar field. We will also require that the EFT approximation holds up to BBN, so the scalar field is frozen before BBN. This way, all the modifications are negligible in the early universe. The scale factor on the onset of the resonance thus is $a_\star<10^{-8}$, and only after that time, the modifications in the gravity becomes relevant.

The resonance frequency $f\sim\omega$ is related to the mass $m$ or the coupling of the quartic interaction $\lambda$ via Eq.~\eqref{eq_3.29} ($\omega\sim ma_t$), which in terms of the model parameter is given by:
\begin{equation}
f\sim\omega\sim 10^{11}\left(\frac{m}{\mathrm{eV}}\right)\left(\frac{a_t}{10^{-4}}\right)\,\mathrm{Hz}.
\end{equation}

For the parameters inside LISA band ($f_{\mathrm{LISA}}\sim 10^{-2}\,\mathrm{Hz}$), we can consider the following relations for the parameters:
\begin{align*}
        m&\sim 10^{-13}\left(\frac{10^{-4}}{a_t}\right)\,\mathrm{eV}, \\
        \beta_\star&\sim 10^{-7}\left(\frac{T_\star}{10\,\mathrm{MeV}}\right)^2\left(\frac{a_\star}{10^{-8}}\right), \\
        \frac{\alpha_\star^4}{\beta_\star}&\sim 10^9\left(\frac{10^{-4}}{a_t}\right)\left(\frac{10\,\mathrm{MeV}}{T_\star}\right)\left(\frac{T_{\mathrm{eq}}}{T_t}\right)^5\left(\frac{H_\star}{\Lambda}\right)^8.
\end{align*}

In order to estimate the gravitational wave spectrum nowadays $\Omega_{\mathrm{GW}}(t_0)$, we will consider the amplification occurring on very short timescales in comparison to the Hubble parameter. Afterwards, the tensor modes will evolve like in GR. This way, the amplification can be estimated simply by using the following equation \cite{Cai:2023ykr}:
\begin{equation} \label{eq_3.13}
\Omega_{\mathrm{GW}}(t_0)=\Omega_{\mathrm{GW}}^{Background}(t_0) \cdot v_p^2,
\end{equation}
where $\Omega^{\mathrm{Background}}_{\mathrm{GW}}(t_0)$ is the spectrum for the background evolution of the universe without considering the resonance, and $v_p$ is the value of the tensor modes after they reach a plateau. As we have seen in Sec.~\ref{sec:2}, the background equation for the GW spectrum can have contributions to the tilt from a non standard early universe dynamics, where the spectrum is given during the radiation dominated era by Eq.~\eqref{eq_2.15}. 

It is important to note that the resonance will cause a direct amplification on certain narrow frequency bands, as we will see in Sec.~\ref{sec:4}. The exact effect in the background is given by numerically solving Eq.~\eqref{eq_3.9}.

\section{Results and Discussion} \label{sec:4}

In this section, we present the relevant plots and assess the detection prospects for the primordial stochastic gravitational wave background, using the SSR effect to probe the stochastic signal. We consider a primordial spectrum parametrized by different values of the tensor-to-scalar ratio $r$ and the tensor spectral index $n_t$, as described in Section~\ref{sec:2}. We consider the spectrum of the modes that re-enter the horizon during the radiation-dominated era. The end of inflation corresponds to higher frequencies, not probed by LISA.

The resonance is obtained by solving Eq.~\eqref{eq_3.9}. To give an example, we will select the resonance parameters to the same values previously provided in \cite{Cai:2023ykr}: $\alpha_\star$, $\beta_\star$ and $\omega$.
For all the plots in this section, we have tuned the value of $\alpha_\star$, which is the parameter that controls the strength of the resonance, to be $\alpha_\star = 0.1$. We also set $\beta_\star$, the parameter that controls how long the modes stay in the resonance band, to the value $\beta_\star = 0.0004$ \footnote{{To ensure that the resonance is detectable by LISA and that the effective field theory description of DHOST gravity does not break down, we require $\alpha_* \leq 0.1$ and $\beta_* \simeq \alpha_*^4$. Additionally, the corresponding mass of the ultralight dark matter (ULDM) must satisfy $m \geq 10^{-13}\mathrm{eV}$.}  We tested several other parameter combinations respecting the relations presented in Sec.~\ref{sec:3} and found that varying $\alpha_\star$ and $\beta_\star$ did not significantly alter the resonance's overall shape.}. Furthermore, we choose $\omega = 10^{-2} \mathrm{Hz}$, so that the resulting resonance peaks fall within the LISA frequency band. This parameter controls when the resonance takes effect, and by changing it, we are able to access when the SSR effect becomes active (by making it higher, we can consider resonances happening in earlier times). The astrophysical foreground, {as shown in Sec.~\ref{sec:Astro}}, is also present in the figures of this section. 

\begin{figure}[t]
    \centering
    \begin{subfigure}{0.48\textwidth}
        \centering
        \includegraphics[width=\linewidth]{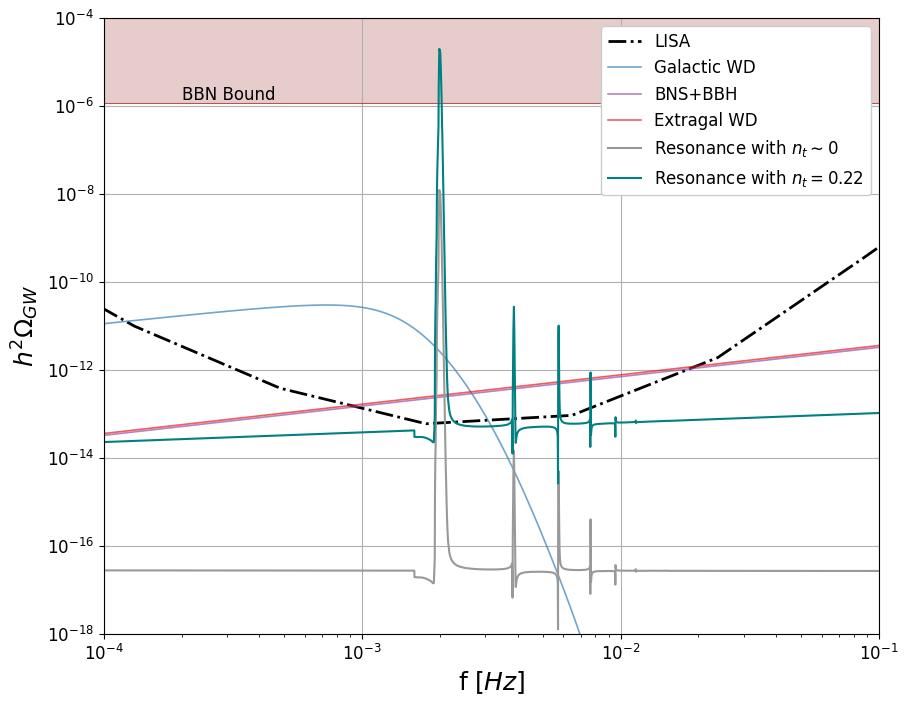}
        \label{fig:r=0.035-a}
    \end{subfigure}
    \hfill
    \begin{subfigure}{0.48\textwidth}
        \centering
        \includegraphics[width=\linewidth]{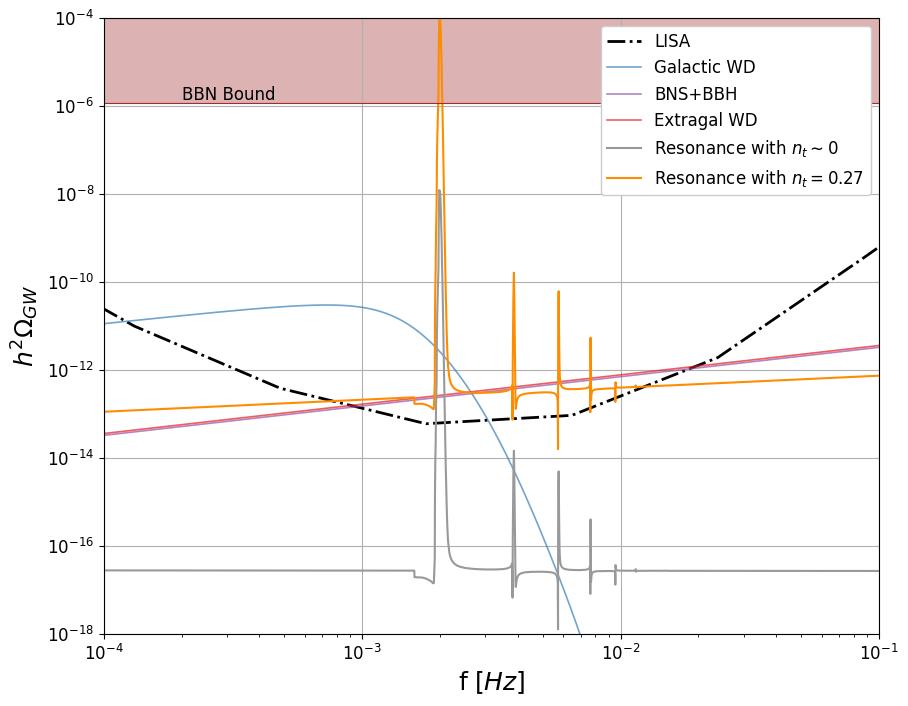}
        \label{fig:r=0.035-b}
    \end{subfigure}
    \caption{Comparison of the resonant GW spectrum with $r = 0.035$ with the astrophysical foreground generated by the unresolved galactic white dwarf (blue); extragalactic white dwarf (red), the sum of the unresolved contribution to binary neutron star, binary black holes and black hole - neutron star binaries (magenta). In the left panel, the teal curve shows the resonance effect on a blue-tilted background with $n_t = 0.22$. In the right panel, the orange curve represents the resonance on a background with {$n_t \sim 0.27$}. The gray curve shows the resonance effect for a standard slow-roll inflationary scenario (where the spectral index follow the inflationary slow-roll consistency relation $n_t \sim 0$). {The red shaded region represents the BBN constraint. This band is shown only as an approximate visual reference to the BBN scale; the actual BBN constraint is imposed through the integrated bound in the original Eq.~\eqref{eq_2.17}, not through the peak amplitude alone.}}
    \label{fig:r=0.035-full}
\end{figure}

In Fig.~\ref{fig:r=0.035-full}, we plot the spectrum using the value provided by Planck for the tensor-to-scalar ratio ($r=0.035$). The standard signal generated by slow roll inflation is shown in light gray, and the signal generated by blue tilted model with a non standard value of $n_t$ in gold. The sensitivity curve of LISA is plotted by using the package SGWBProbe, according to the procedure described in Ref.~\cite{Campeti:2020xwn}. {Notice that the BBN bound is imposed through the integrated quantity in Eq.~(2.20), and not through the peak value of $\Omega_{\rm GW}(f)$ alone. Therefore, a resonant spectrum is not excluded simply because its peak rises above a horizontal reference level in the plot. The relevant criterion is the total area under the spectrum. The horizontal BBN line shown in the figures should thus be understood only as an approximate visual guide.}


In the left panel of Fig.~\ref{fig:r=0.035-full}, we have the background model with highest tensor spectral index ($n_t \sim 0.22$) that does not produce a detectable signal without the resonance in LISA band. The first peak that already had sufficient amplitude to be detected by LISA is amplified due to the blue tilted background spectrum, which enhance modes at higher frequencies. With this choice of parameter, the other peaks, which would not be detectable in the standard background, are amplified to higher amplitudes, bringing the signal into observable range above all the astrophysical foreground.

In the right panel of Fig.~\ref{fig:r=0.035-full}, we show the background model with highest value of the tensor spectral index {($n_t \sim 0.27$)} that fully avoids the constraints from BBN
\footnote{{The resonant feature develops over a  timescale related to the inverse frequency of the resonant oscillations. The BBN bound should, strictly speaking, be applied to the portion of the present-day spectrum with $f \ge f_{\rm BBN}$. Therefore, for the resonance frequency considered here, the amplitude of the  features at BBN can be smaller than the fully developed value shown in our spectra. In this work, by imposing the BBN bound using the fully developed resonant spectrum, we adopt the most stringent, and hence conservative, implementation of this constraint.}}.
{With this value for the spectral index, the background remains with a smaller -but comparable- amplitude than the astrophysical foreground across all of LISA's frequency range. The resonance effect could still be used to probe the background parameters by the signal obtained by the peaks (which are all above the astrophysical foreground).}


\begin{figure}[t]
    \centering
    \includegraphics[width=0.6\linewidth]{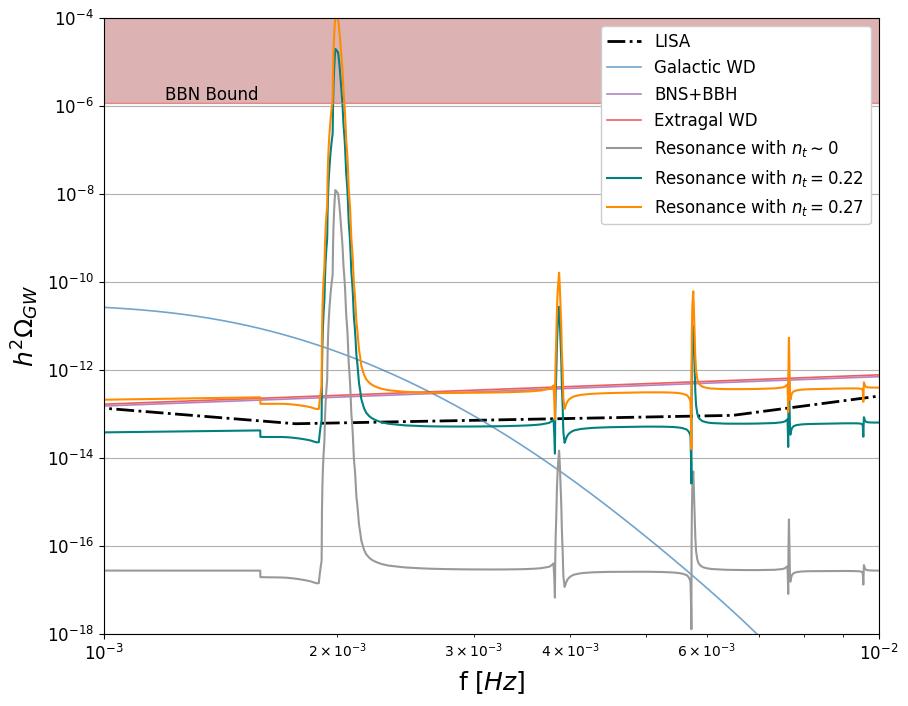}
    \caption{Plot comparing all the three models. In this plot, the background has $r = 0.035$. We show the case with $n_t \sim 0$ in gray, the blue tilted case with {$n_t \sim 0.27$} in orange and $n_t = 0.22$ in teal. The legend is the same as in Figure~\ref{fig:r=0.035-full}.}
    \label{fig:r=0.035 - comparison}
\end{figure}

In Fig.~\ref{fig:r=0.035 - comparison}, we show a close up view of all the three models together with the astrophysical foreground for a more direct visual comparison of the effects generated by the difference in the value of $n_t$. It is worth emphasizing that all the models considered here are consistent with the upper bounds on the amplitude of the SGWB imposed by LIGO.

\begin{figure}[t]
    \centering
    \begin{subfigure}{0.48\textwidth}
        \centering
        \includegraphics[width=\linewidth]{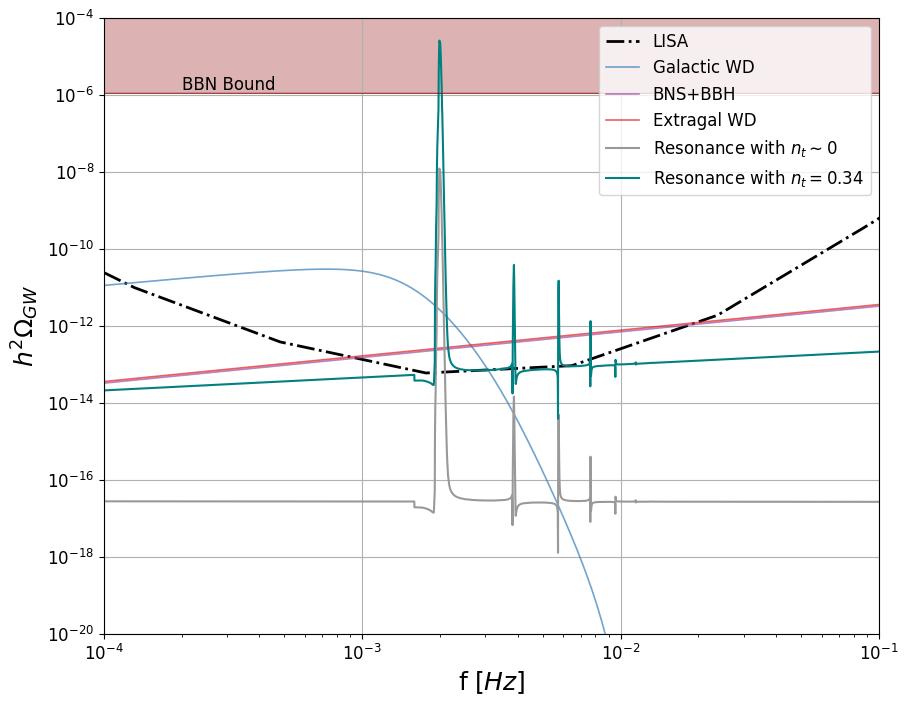}
        \label{fig:r=0.001-a}
    \end{subfigure}
    \hfill
    \begin{subfigure}{0.48\textwidth}
        \centering
        \includegraphics[width=\linewidth]{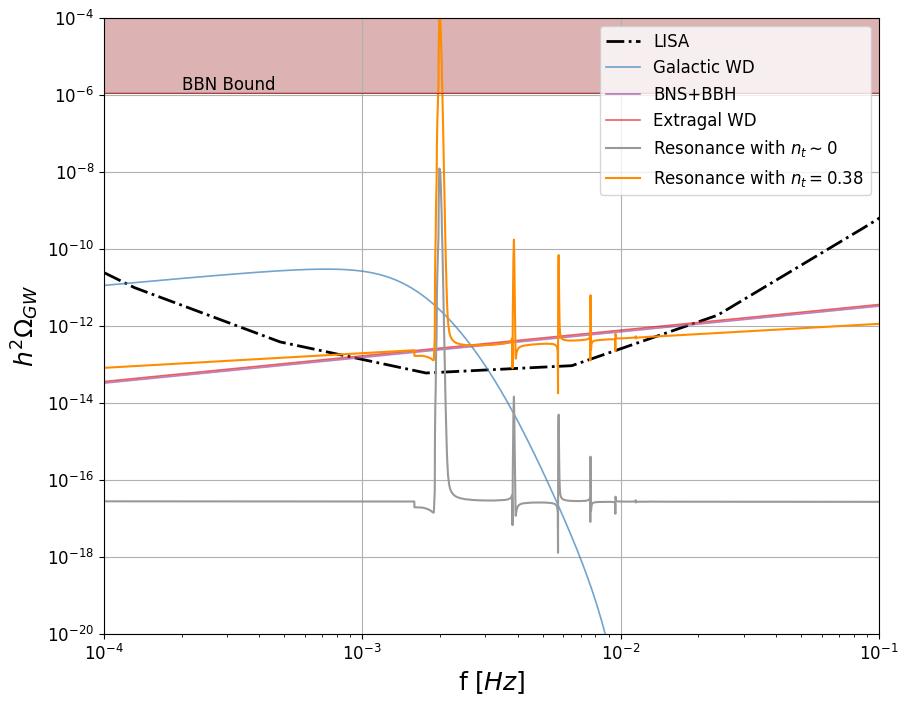}
        \label{fig:r=0.001-b}
    \end{subfigure}
    \caption{Comparison of GW spectra for the resonance effect with a background parametrized by $r=1 \cdot 10^{-3}$. The colors are the same as in Fig.~\ref{fig:r=0.035-full}. In this case, the curve represents the same resonance effect happening in a blue-tilted background with $n_t = 0.336$ (teal) in the left panel and with {$n_t \sim 0.381$} (orange) in the right panel. The red shaded region represents the excluded region by the BBN constraint and the astrophysical foreground is the same as before.}
    \label{fig:r=0.001-full}
\end{figure}

On another note, the resonance effect also allows us to explore what happens to the detection of a background with lower values of the tensor-to-scalar ratio. In the case of $r = 1 \cdot 10^{-3}$, as shown in Fig.~\ref{fig:r=0.001-full}, we can see that the first peak would still be detectable in the baseline model. However, if we compare the resonance to a model with a maximum value of the tensor index $n_t = 0.336$ (or {$n_t \sim 0.381$} to satisfy the limit imposed by the BBN constraint), {the resonance peaks} would be inside the LISA's detectable frequency band, similar to the previous case of $r=0.035$. {Thus, for values of the spectral index between \(n_t \in [0.336,0.381]\), the background would be smaller than the astrophysical contribution, making it difficult to probe it without the resonance.}

\subsection{The spectrum of SGWB in the presence of resonance: The role of each parameter}

\begin{figure} [t]
    \centering
    \includegraphics[width=0.6\linewidth]{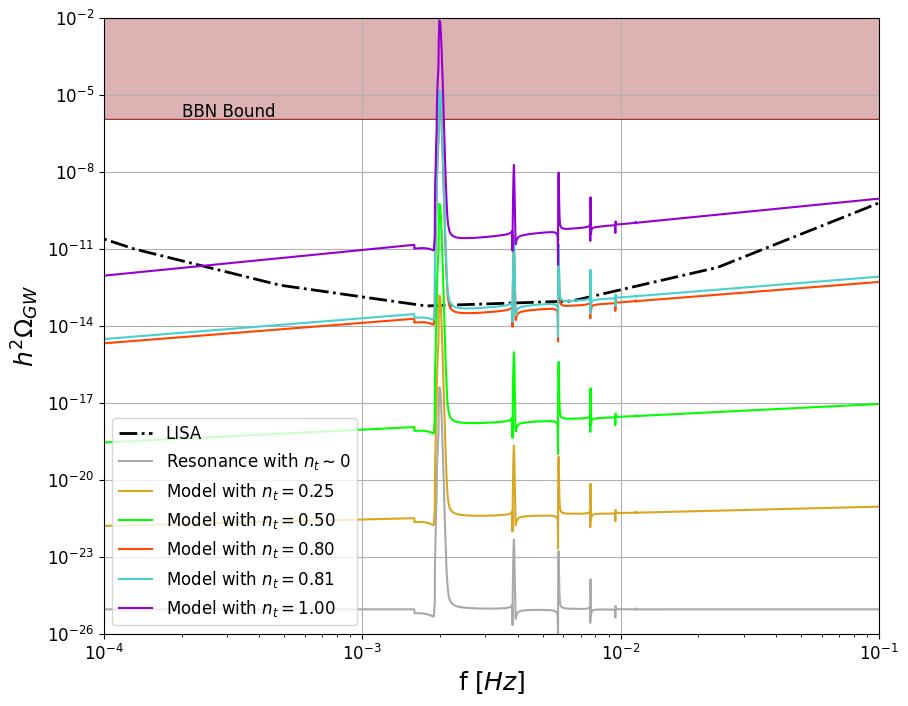}
    \caption{Comparison of different values of tensor spectral index in the range $n_t \in [0,1]$. In this plot, we consider $r = 10^{-10}$. Specifically, we show the curves where {$n_t = 0.80$ (in red)}, which would be able to satisfy the BBN bound and the case with {$n_t=0.81$ (in blue)}, which the background signal would be inside LISA's sensitivity but is ruled out by the BBN constraint. LISA's sensitivity curve is in black. \color{red}{The light red shaded region represents approximate visual reference for the BBN constraints.}}
    \label{fig:r=1e-10 - multi}
\end{figure}

The most promising detection prospects for a blue tilted spectrum happens if the value of $r$ is orders of magnitude lower than the Planck upper bound. For a better visualization of this effect, we plot the results for $r = 10^{-10}$ in Fig.~\ref{fig:r=1e-10 - multi}, which allows us to explore a higher range of blue tilted spectrum. This figure shows how the signal changes different values of $n_t$, while it lies in the range $n_t \in [0,1]$. Changing the value of the spectral index generates a stronger amplification for the smaller peaks as compared to the first one, leading them to a possible observable range for an value of $n_t$ that could be more favorable to blue tilted models that are studied in the literature.
Overall, the effect of a lower value of tensor-to-scalar ratio is a considerably reduction of the amplitude of the full spectrum, across every frequency. In that sense, it becomes difficult to determine whether the reduction is due to any alteration in the SSR mechanism (e.g., a reduction of the overall amplitude of the resonance) or simply a lower value of $r$ of the background spectrum by just studying the signal by itself.

\begin{figure} [t]
    \centering
    \includegraphics[width=0.6\linewidth]{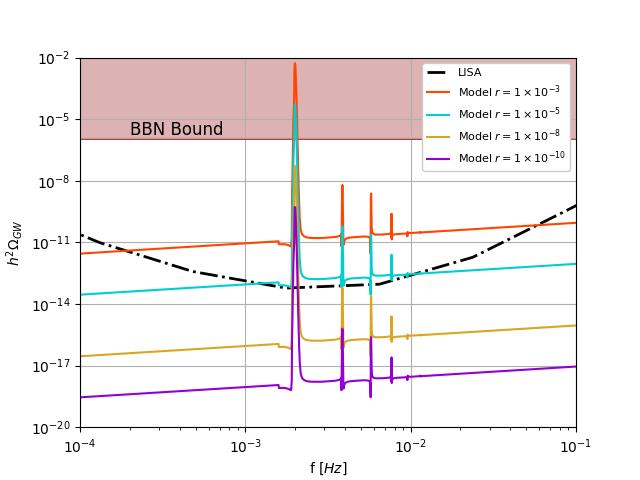}
    \caption{Plot comparing different values of $r$ when $n_t = 0.5$. The blue curve represent the case of $r=0.035$ with the normal tensor tilt $n_t \sim 0$, while the other curves represents the cases of $r = 10^{-10}$ (magenta), $r = 10^{-8}$ (orange), $r = 10^{-5}$ (blue) and $r = 10^{-3}$ (red). The BBN bound is represented by the red shaded region.}
    \label{fig:n_t=0.5 - multi}
\end{figure}

Figure~\ref{fig:n_t=0.5 - multi} shows the effect of the change in $r$ in the theoretical prediction for the spectrum. In this plot, we have fixed the value of the tensor index to be $n_t = 0.5$ and plotted the prediction for the resonant GW spectrum for $r$ to range on four specific values ranging from $r = 10^{-10}$ until $r = 10^{-3}$, comparing with the results for the SSR happening on a background parametrized by the Planck value, $r=0.035$ and $n_t \sim 0$. As expected, the only effect that the tensor-to-scalar ratio has in the signal is to change the total amplitude of the spectrum. In order for the resonant spectrum to be detectable, a lower value of $r$ requires a correspondingly higher blue tilted spectral index for the primordial background to compensate for the reduced amplitude.

\subsection{Probing the primordial spectrum parameters with resonance}

As presented in the previous section, one of the interesting features of the resonance effect is the possibility to amplify the cosmological signal in the LISA frequency band. By considering different combinations of the tensor-to-scalar ratio and the spectral index, we find that the resonance can amplify the gravitational-wave background, potentially bringing it within observational reach. Although smaller values of  $r$ relax the BBN constraints for a given spectral index, the presence of resonance can  lead to a violation of these bounds. It is therefore important to determine how resonance modifies the allowed parameter space.

\begin{figure}[t]
    \centering
    \begin{subfigure}{0.48\textwidth}
        \centering
        \includegraphics[width=\linewidth]{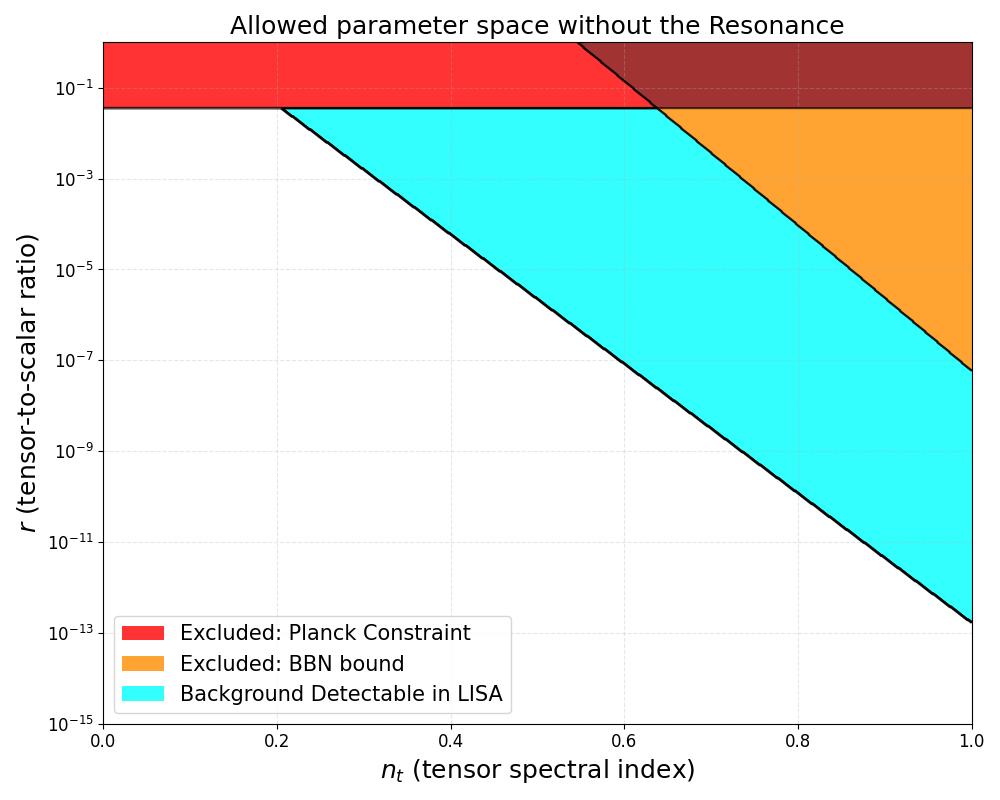}
        \label{fig:n_t/r no resonance}
    \end{subfigure}
    \hfill 
    \begin{subfigure}{0.48\textwidth}
        \centering
        \includegraphics[width=\linewidth]{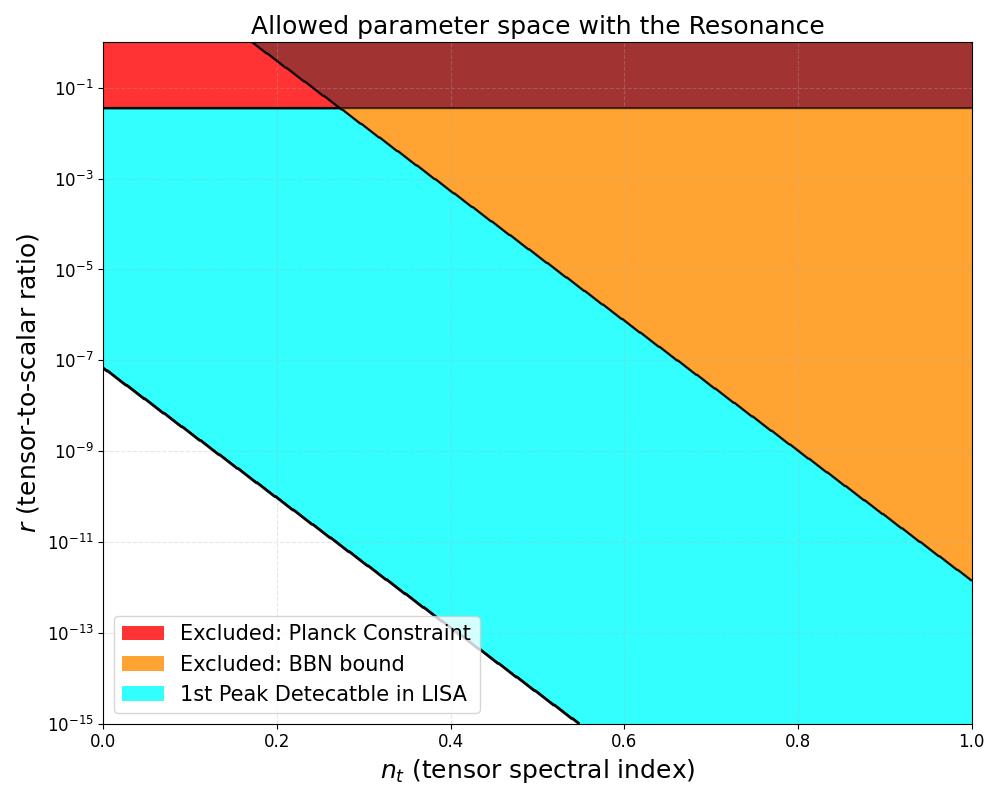}
        \label{fig:n_t/r with resonance}
    \end{subfigure}
    \caption{Comparison of the parameter space in the $r$-$n_t$ plane that would make a detectable signal in LISA (blue region). On the left, we have the constraints in the spectrum without the resonance, and on the right, we have the constraints when the resonance happens. The red region is related to the parameters excluded by Planck, while the orange region represents the parameter subset excluded by the BBN bound.}
    \label{fig:n_t/r_full}
\end{figure}

Considering the minimum frequency at which LISA can detect a signal, $f_{\text{min}} \sim 1.8 \times 10^{-3}  \mathrm{Hz}$, we plot in Figure \ref{fig:n_t/r_full} how the resonance affects the possible parameter space accessible to LISA. In the left panel, we show the allowed parameter space for a cosmological background in the absence of the resonance that could generate a detectable signal, in the blue shaded region. The red region represents the Planck upper bound in $r$ and the orange region denotes the part of the parameter space that violates the BBN bound within the LISA frequency range.

On the right panel of Fig.~\ref{fig:n_t/r_full}, we show the allowed parameter space for a cosmological background in the presence of the resonance studied in this work. As can be seen, the presence of resonance significantly alters the allowed parameter space compared to the case without resonance. The region corresponding to a detectable signal (blue) changes, indicating that the presence of the resonance can amplify a smaller signal to detectable levels or even exceed the BBN limit for a broader range for the value of the tensor-to-scalar ratio for the background, as shown in Figure \ref{fig:r=1e-10 - multi}. On another hand, for a fixed value of $r$, the presence of resonance allows to probe a wider range of the spectral index, as illustrated in Figure \ref{fig:n_t=0.5 - multi}.  With the resonance included, we observe a shift in the region that violates the BBN constraint when analyzing the first peak, which occurs at a frequency $f_{\text{peak 1}} \sim 2.0 \times 10^{-3} , \mathrm{Hz}$.

\begin{figure} [t]
    \centering
    \includegraphics[width=0.6\linewidth]{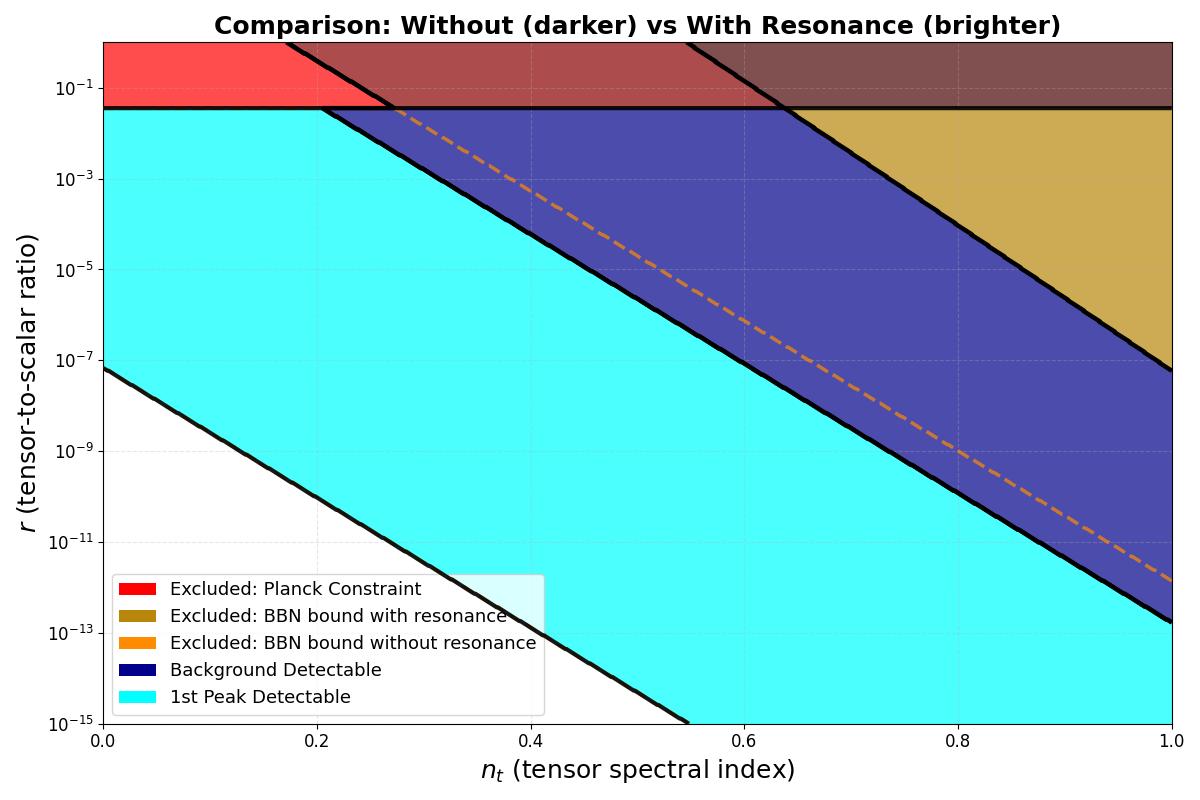}
    \caption{Comparison  of the parameter space in the $r$-$n_t$ plane that could be probed by LISA in the absence (dark blue region) and in the presence (light blue region) of the resonant peaks. The red region is related to the parameters excluded by Planck, while the orange region represents the parameter subset excluded by the BBN bound. {In the case of the resonance, the whole region above the orange dashed line is excluded by the resonance.}}
    \label{fig:n_t/r comparison}
\end{figure}

In Figure \ref{fig:n_t/r comparison}, we compare this change in the parameter space visually. The dark blue region is related to the parameter space that generate a detectable signal in LISA with the absence of the resonance, while the light blue is related for the parameter space in the presence of it. As one can see, the region that violates BBN constraint shifts when one consider the resonance {(from the dark orange to the full region above the orange dashed line)}. 

\section{Conclusion}
It has been shown in the literature that extensions of general relativity involving additional gravitational degrees of freedom, such as scalar fields, can modify the propagation of gravitational waves, leading to the appearance of resonant features in the gravitational wave spectrum. We investigated the prospects for detecting a primordial GW signal in the presence of such resonance. While the ultralight dark matter scenario considered in this paper is only one realization of the parametric resonance effect, our overall qualitative results remains general.

We showed that, in certain scenarios, the resonance can enhance the spectrum to amplitudes within the sensitivity range of the forthcoming detector LISA, leading to a possible detection of one or more features of the resonant amplification. Furthermore, we have demonstrated how the presence of resonance significantly reshapes the region of parameter space accessible to LISA in the $r$-$n_T$ plane, enabling the experiment to probe lower primordial signals than would be possible in the absence of resonance.

As is well known, within the narrow frequency band probed by a single experiment there are limited prospects for independently constraining the parameters $r$ and $n_t$, even in the idealized case where a primordial signal is unambiguously detected. The restricted frequency coverage leads to a strong degeneracy between these parameters. To obtain independent constraints on $n_t$, joint measurements of the spectrum across multiple frequency bands would be required, enabling a multi-frequency test of the gravitational-wave spectrum. Encouragingly, the coming years offer promising prospects for such multi-frequency observations, which could help break the degeneracy in the $r$-$n_t$ plane. However, for more resonant peaks to be detectable, it would be necessary to have a blue tilted spectrum, which would be a hint for new physics.

\acknowledgments
We are grateful to Manqi Li and Mian Zhu for valuable discussions and comments. This work was supported in part by the National Key R\&D Program of China (2021YFC2203100), by NSFC (12433002, W2533006), by CAS young interdisciplinary innovation team (JCTD-2022-20), by 111 Project (B23042), by CSC Innovation Talent Funds, by USTC Fellowship for International Cooperation, and by USTC Research Funds of the Double First-Class Initiative. 
I.O.C.P. is also supported by ANSO and CAPES. L.G is supported by research grants from Conselho Nacional de Desenvolvimento Cientıfico e Tecnologico (CNPq), Grant No. 307636/2023-2 and from the Fundacao Carlos Chagas Filho de Amparo a Pesquisa do Estado do Rio de Janeiro (FAPERJ), Grant No. E-26/204.598/2024. Z.W the is supported in part by the Basic
and Frontier Research Project of PCL (grant No. 2025QYB012) and the Major Key project of Peng Cheng Laboratory.


\bibliography{bibliography}

\end{document}